\documentclass[aps,prl,twocolumn,superscriptaddress,showpacs]{revtex4-1}
\usepackage{amsmath,amssymb,graphicx}

\usepackage[utf8]{inputenc}
\usepackage[T1]{fontenc}
\usepackage{xcolor}

\IfFileExists{newtxtext.sty}
   {\usepackage{newtxtext,newtxmath}}
   {\IfFileExists{stix.sty}
      {\usepackage{stix}}
      {\IfFileExists{mathptmx.sty}
      {\usepackage{mathptmx}}{} } }

\usepackage{textcomp}

\usepackage{bm}

\IfFileExists{siunitx.sty}{\usepackage{booktabs,siunitx}}{}

\pdfoutput=1
\usepackage{color}
\definecolor{LinkColor}{rgb}{0.256,0.439,0.588}
\usepackage{hyperref}
\hypersetup{
   pdfauthor={Xiao Yan Xu, K. S. D. Beach, Kai Sun, F. F. Assaad, and Zi Yang Meng},
   pdftitle={Topological phase transition with SO(4) symmetry in (2+1)d interacting Dirac fermions},
   colorlinks=true,
   citecolor=LinkColor,
   linkcolor=LinkColor,
   urlcolor=LinkColor
}

\DeclareMathOperator{\sgn}{sgn}
\renewcommand{\vec}[1]{\mathbf{#1}}

\usepackage{pifont}

\newcommand{\circone}{\raisebox{-0.085em}{\ding{192}}}
\newcommand{\circtwo}{\raisebox{-0.085em}{\ding{193}}}
\newcommand{\circthree}{\raisebox{-0.085em}{\ding{194}}}
\newcommand{\circfour}{\raisebox{-0.085em}{\ding{195}}}
\newcommand{\circfive}{\raisebox{-0.085em}{\ding{196}}}

\begin{document}

\title{Topological phase transitions with SO(4) symmetry in (2+1)d interacting Dirac fermions}

\author{Xiao Yan Xu}
\affiliation{Beijing National Laboratory for Condensed Matter Physics and Institute of Physics, Chinese Academy of Sciences, Beijing 100190, China}
\author{K. S. D. Beach}
\affiliation{Department of Physics and Astronomy, The University of Mississippi, University, Mississippi 38677, USA}
\author{Kai Sun}
\affiliation{Department of Physics, University of Michigan, Ann Arbor, MI 48109, USA}
\author{F. F. Assaad}
\affiliation{Institut f\"ur Theoretische Physik und Astrophysik, Universit\"at W\"urzburg, 97074 W\"urzburg, Germany}
\author{Zi Yang Meng}
\affiliation{Beijing National Laboratory for Condensed Matter Physics and Institute of Physics, Chinese Academy of Sciences, Beijing 100190, China}

\date{\today}

\begin{abstract}
Interaction-driven topological phase transitions in Dirac semimetals are investigated by means of large-scale quantum Monte Carlo (QMC) simulations. The interaction among Dirac fermions is introduced by coupling them to Ising spins that realize the quantum dynamics of the two-dimensional transverse field Ising model. The ground state phase diagram, in which the tuning parameters are the transverse field and the coupling between fermion and Ising spins, is determined. At weak and intermediate coupling, a second-order Ising quantum phase transition and a first-order topological phase transition between two topologically distinct Dirac semimetals are observed. Interestingly, at the latter, the Dirac points smear out to form nodal lines in the Brillouin zone, and collective bosonic fluctuations with SO(4) symmetry are strongly enhanced. At strong coupling, these two phase boundaries merge into a first-order transition. 
\end{abstract}


\maketitle

Dirac fermions in (2+1)d emerge in a number of solid state systems such as graphene~\cite{Novoselov2004}, surface states of 3D topological insulators~\cite{Xia2009} and d-wave superconductors~\cite{Vojta00,Kim08,Huh2008}. A parallel research track involves interaction-driven topological phase transitions in (2+1)d systems~\cite{Sun2009,Varney2010,Varney2011,Hohenadler10,Hohenadler2012,Lang2013,You2016,Wu2016,He2016b,He2016c,Slagle2015,He2016a}, which exhibit exotic quantum critical points. In this context, an important question is what new physics would emerge if Dirac fermions and topological phase transitions were brought together via electronic interactions?

Timely developments in quantum Monte Carlo (QMC) techniques offer an opportunity to address this question: rather than simulating an explicit interaction between fermions, one can instead introduce bosonic fields that mediate fermion interaction. The key insight is that the form of these fields need not be limited to what would arise from a Hubbard-Stratonovich decomposition of explicit fermion interactions. One interesting direction is to endow the bosonic degrees of freedom with quantum dynamics of their own, such that they can be tuned through a quantum critical point (QCP). Examples of this approach include QMC studies of fermions in 2d coupled to nematic~\cite{Schattner2015a,Lederer2016}, antiferromagnetic~\cite{Berg12,Schattner2015b,Gerlach2016},  or Ising gauge fluctuations \cite{Assaad16,Gazit16}, which have revealed interesting features of  metallic QCP as well as realizations of deconfined phases.

\begin{figure}[htp!]
\includegraphics{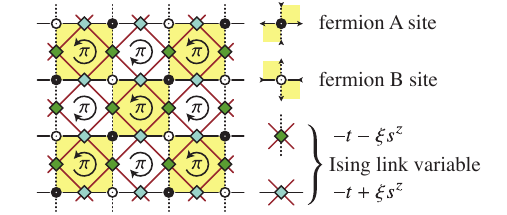}
\caption{Fermions on a square lattice coupled to a transverse-field Ising model. Ising spins live on the links between adjacent fermionic sites and modify the corresponding hopping integral. The solid (dotted) lines indicate the positive (negative) character of the coupling between the Ising spins and the hopping. A magnetic flux of magnitude $\pi$ runs through each plaquette. The arrows indicate the direction of the counter-circulating currents.}
\label{fig:model}
\end{figure}

In this paper, we consider a system of Dirac fermions on the square lattice, coupled to Ising spins that decorate the nearest-neighbor links. Ordering of the Ising spins breaks the underlying $C_{4v}$ symmetry and allows for anisotropic velocity renormalization of the Dirac points. We discover a first-order, interaction-driven topological phase transition, at which collective bosonic fluctuations with SO(4) symmetry comprised of Dirac fermions manifest. Across the transition, the bulk topological index, associated with the topological Dirac semimetal, flips its value and thereby shifts the momentum of the topologically protected edge states in the projected 1D Brillouin zone (BZ). Upon further increase of the transverse field and/or coupling, the second-order phase boundary of the Ising ordering merges with the topological phase transition into a first-order transition line.

{\it Model and Method}\,---\,We use sign-problem-free projector QMC~\cite{Sugiyama86,Sorella89,AssaadEvertz2008}
to simulate a model of fermions on the $\pi$-flux square lattice in which (i) the hopping is mediated by Ising spins positioned on the nearest-neighbor links and (ii) the Ising spins interact via a two-dimensional transverse field Ising model (TFIM) on the dual lattice. The time-reversal symmetry of Eq.~\eqref{EQ:hamiltonian} guarantees positivity of the fermion determinant. The detailed description of QMC implementation, including local-plus-global updates, is given in the supplemental material (SM)~\cite{suppl}. The model, illustrated in Fig.~\ref{fig:model}, has a Hamiltonian
\begin{equation}\label{EQ:hamiltonian}
\begin{split}
H &=\sum_{j,\delta}\bigl(-t+\sgn(\delta) \xi s_{j,j+\delta}^{z}\bigl)\bigl(c^{\dagger}_{j}c^{}_{j+\delta}\ e^{i\frac{\pi}{4}\sgn(\delta)}+\text{h.c.}\bigr)\\
&\qquad\qquad-J\sum_{\langle bb'\rangle}s_{b}^{z}s_{b'}^{z}-h\sum_{b}s_{b}^x.
\end{split}
\end{equation}

\begin{figure}[!htp]
\includegraphics{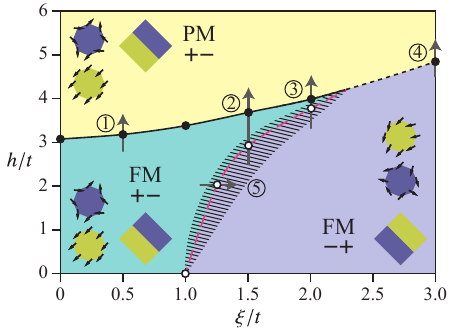}
\caption{The $\xi$--$h$ phase diagram. FM and PM indicate the ferromagnetic and paramagnetic Ising spin states.
The color inside circles with black arrows relates to topological index: the windings of the 2D vectors $\mathbf{n}$ along $\phi_1=\pm \pi/2$ loop for $h/t=2,\xi/t=0.2$ inside the left FM phase; $h/t=4,\xi/t=1.5$ inside PM phase; $h/t=4, \xi/t=3$ inside right FM phase. The sqaure insets stand for the 2d BZ with purple area topologically nontrivial (winding number  $\mathcal{W}=-1$) while green area trivial (winding number $\mathcal{W}=0$).  Across the red dashed line, where the topological phase transition happens, the two areas of the BZ switch. The cross-hatched area highlights the enhancement of collective SO(4) bosonic fluctuations in the fermion sector. The black solid line is the FM-to-PM magnetic phase transition in Ising spins. Along paths \protect\circone, \protect\circtwo, and \protect\circthree, this transition is continuous, along paths \protect\circfour\ and \protect\circfive\ it becomes first order. }
\label{fig:phase-diagram}
\end{figure}

Here, we have spin-1/2 fermions $c_j^\dagger=\bigl(c_{j\uparrow}^\dagger\ ,\ c_{j\downarrow}^\dagger\bigr)$ on the each site $j$ of a square lattice and an Ising spin $s_b^z= \pm 1$ on each bond with $b$ being the bond index.
The dynamics of Ising variables is governed by a ferromagnetic TFIM. For the fermions, the nearest-neighbor hopping with phase factor $\frac{\pi}{4}\sgn(\delta)$ generates a $\pi$ flux through each plaquette.
Because $\sgn(\delta)$ distinguishes the horizontal ($+$)  and vertical ($-$) bonds, the spins modify the hoppings oppositely along the $x$ and $y$ directions with coupling strength $\xi$. This spin-fermion coupling connects the $Z_2$ Ising symmetry to the $\pi/2$ space rotational symmetry,
i.e., an Ising ferromagnetic order will induce hopping anisotropy between $x$ and $y$ and hence an electronic nematic ordering \cite{Kivelson1998}.

In addition to the Ising and lattice point group symmetries, the fermionic degrees of freedom also preserve internal symmetries,
independent of the ordering pattern of the Ising fields. $\text{SU}_S(2)$ spin rotations, with generators $S^{\alpha}$, and the particle-hole transformation  $P^{-1}c^\dagger_{i,\downarrow} P = (-1)^{i}c^{}_{i,\downarrow}$ leave the Hamiltonian invariant. One can also define operators $\eta^{\alpha} = P^{-1} S^{\alpha} P$  that obey $\bigl[ \eta^{\alpha}, S^{\beta}\bigr] = 0$. Hence, the full symmetry of the model is $\text{SU}_S(2) \otimes \text{SU}_\eta(2) \otimes \mathbb{Z}_{2,\text{ph}}$, corresponding to the SO(4) symmetry of Ref.~\onlinecite{Yang1990}. As a consequence of this enhanced symmetry, antiferromagnetic and superconducting states may coexist \cite{Feldbacher2003}.

\begin{figure}
\includegraphics{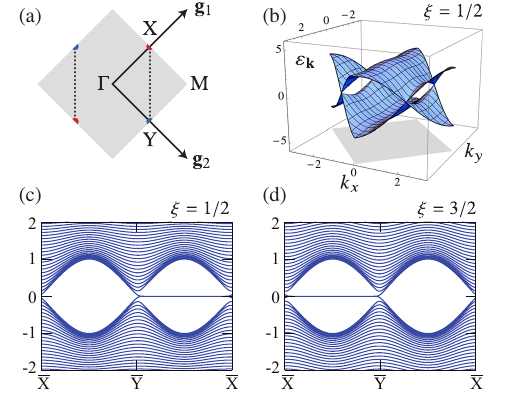}
\caption{(a) The BZ of the $\pi$-flux square lattice is constructed from the reciprocal lattice vectors $\vec{g}_1 = (\pi,\pi)$ and $\vec{g}_2 = (\pi,-\pi)$. High-symmetry points in the BZ are labeled by $\Gamma$, M, X, and Y. Two Dirac points are located at $\text{X} = (\pi/2,\pi/2)$ and $\text{Y} = (\pi/2,-\pi/2)$. (b) The electronic dispersion $\varepsilon_\vec{k}$ at zero transverse field ($h=0$) for coupling strengths $\xi=1/2$. (c) and (d) Spectra with open boundaries plotted in the projected BZ $\bar{X}-\bar{Y}-\bar{X}$ at (c) $\xi=1/2$ and (d) $\xi=3/2$. The states at zero energy are topologically protected edge states.}
\label{fig:dispersion}
\end{figure}

{\it Results}\,---\,We set $J=t=1$ in the calculation. The QMC results are summarized in the $\xi$--$h$ phase diagram of Fig.~\ref{fig:phase-diagram}. At $\xi=0$, the fermions and the Ising spins decouple. Because of the $\pi$ flux,
the fermions form a Dirac semimetal with Dirac points located at the $X$ and $Y$ points of the BZ.
For the spin degrees of freedom, a 2D TFIM is recovered, where paramagnetic ($\langle  s_b^z \rangle=0$) and ferromagnetic ($\langle s_b^z  \rangle \ne 0$) phases are separated by a quantum critical point at $h_c=3.04(2)$, consistent with literature~\cite{Pfeuty1971}.

As we increase $\xi$, because the coupling between Dirac fermions and Ising anisotropy is perturbatively irrelevant~\cite{Vojta00, Vafek02, Balents1998} \footnote{Because of symmetry, the Ising ordering in our model only modifies the Fermi velocity without changing the locations of the Dirac points, which implies that in the long-wavelength limit, the coupling between the Ising fields and fermion bilinears requires at least one spatial derivative and thus is irrelevant in strong analogy to case (C) in Ref.~\onlinecite{Vojta00}.}, the second-order Ising phase transition remains but with a renormalized critical $h_c$. This second-order phase boundary is indicated in Fig.~\ref{fig:phase-diagram} as the solid black line. Upon further increase of $\xi$, new phenomena beyond the TFIM arise.  At $\xi>1$, a new phase boundary emerges (red dashed line), around which strong fluctuations in a fermion bilinear are observed (the shaded area). As will be shown below, this new phase boundary is a first-order topological phase transition. At strong coupling ($\xi>2.2$), the topological phase boundary and the second-order Ising phase boundary merge together into a single first-order phase boundary (black dashed line). 

To understand the topological phase transition, we first focus on the exactly solvable limit at $h=0$ (the horizontal axis of the phase diagram). In this limit, the Ising spins $s_b^z$ have no quantum fluctuations and must choose a classical ferromagnetic spin configuration, with $s_b^z=+1$ or $s_b^z=-1$; the system then reduces into a free-fermion problem.
At finite $\xi$, as discussed above, the Ising ordering reduces the four-fold rotational symmetry down to two-fold ($C_{4v}$ to $C_{2v}$).
However, as long as $\xi\ne\pm1$, the system remains a Dirac semimetal and the location of the Dirac points are pinned to $X$ and $Y$
by the $C_{2v}$ symmetry. The reduction of the rotational symmetry is reflected by the anisotropy in Fermi velocity.
Near a Dirac point (e.g. $X$), the linearized Hamiltonian is $H_X=2(t-\xi) k_x\sigma_x+2(t+\xi) k_y\sigma_y$, i.e., the Dirac points have different Fermi velocity along $x$ and $y$, as shown in Fig.~\ref{fig:dispersion}(b).  At $\xi=1$, the velocity in one direction vanishes, and a nested Fermi surface develops. This metallic state marks a topological phase transition between two topologically distinct Dirac semimetals.

To define the band topology for this 2D Dirac semimetal, we utilize the idea of dimension reduction,
in analogy to 3D Weyl semimetals \cite{Wan2011,Weng2015,Burkov2011, Hosur2012}. In the momentum space, we can define the Hamiltonian for each momentum point $H(\phi_1,\phi_2)$,
where a momentum point is labeled as $\mathbf{k}=\phi_1 \mathbf{g_1}/2\pi+\phi_2 \mathbf{g_2}/2\pi$ with
$\mathbf{g}_{1}$ and $\mathbf{g}_{2}$ being the reciprocal lattice vectors shown in Fig.~\ref{fig:dispersion}(a), and $-\pi<\phi_1<\pi$ and $-\pi<\phi_2<\pi$. For a fixed $\phi_1$, i.e., along a line in the BZ, the Hamiltonian, as  a function of $\phi_2$, can be treated as 1D system.
For $\xi\ne \pm 1$ and $\phi_1\ne 0$ or $\pi$, it is easy to verify that such a 1D system has a finite energy gap.
Because of the chiral symmetry defined above, this gapped 1D system falls into the AIII class of Refs.~\cite{Kitaev2009, Schnyder2008}  and thus supports an integer-valued
topological index, i.e., a winding number.
As required by the chiral symmetry, the Hamiltonian can be written as
$H(\phi_1,\phi_2)= H_x(\phi_1,\phi_2) \sigma_x+H_y(\phi_1,\phi_2) \sigma_y$, where $\sigma_x$ and $\sigma_y$ are two of the
Pauli matrices. A 2D unit vector can be defined as $\mathbf{n}=\frac{(H_x, H_y)}{\sqrt{H_x^2+H_y^2}}$ for  $\xi\ne \pm 1$ and $\phi_1\ne 0$ or $\pi$.
For a fixed $\phi_1$, as $\phi_2$ increases from $-\pi$ to $+\pi$, this 2D vector winds $\mathcal{W}$ times around the unit circle. The winding
number $\mathcal{W}$ is the topological index of this Dirac semimetal.
For $0<\xi<1$, $\mathcal{W}=-1$ for $0<\phi_1<\pi$, and $\mathcal{W}=0$ for $-\pi<\phi_1<0$. For $\xi>1$, the topological index flips its value to $\mathcal{W}=0$ for $0<\phi_1<\pi$ and $\mathcal{W}=-1$ for $-\pi<\phi_1<0$, i.e., $\xi=1$ is a topological transition.
To change a topological index, the bulk band gap must close, which results in the nodal lines.
The experimental signature for the band topology and topological transition lies in the edge states. As shown in Fig.~\ref{fig:dispersion}(c) and (d), for $0<\xi<1$, the nontrivial bulk topological index results in zero-energy edge states for projected momentum $0<\phi_1<\pi$. For $\xi>1$, however, the zero-energy edge states shift to momentum $-\pi<\phi_1<0$.

For $h\ne 0$, as shown in the SM~\cite{suppl}, the topological index can be defined via an effective Hamiltonian, which is the inverse of the
single-particle fermionic Green's functions at zero frequency~\cite{Wang2012}. Away from the shaded region
in Fig.~\ref{fig:phase-diagram}, the QMC simulations show a finite single-particle gap for momentum points away from the Dirac points
($X$ and $Y$), where the effective Hamiltonian is well-defined. One can then use it to evaluate the topological index, following the same procedure described above.
In the insets of Fig.~\ref{fig:phase-diagram}, we presented the winding number for $\phi_1=\pm \pi/2$ at different values of $\xi$ and $h$. We find that the FM phase has two distinct topological semimetal phases (at small and large $\xi$).
For the PM phase, the band topology coincides with the small $\xi$ FM phase.

{\it Phase transitions}\,---\,To better understand the phase transitions, we explore the phase diagram with several parameter scans. Below, we discuss QMC data along the five exemplary paths labeled \circone\ through \circfive\ in Fig.~\ref{fig:phase-diagram}.

\begin{figure}
\includegraphics[width=0.9\columnwidth]{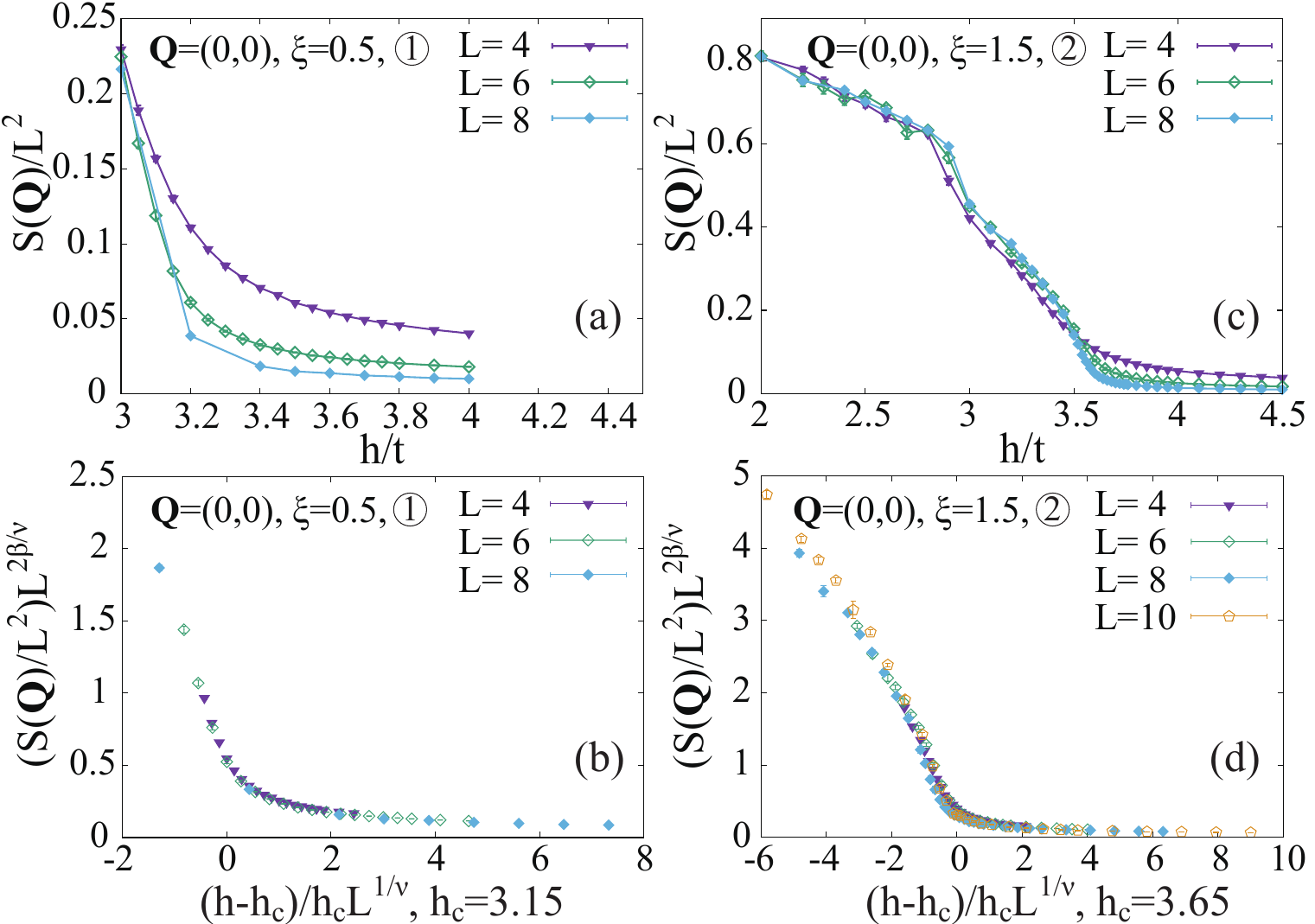}
\caption{(a) Correlation function $S(\vec{Q})$ of Ising spins along path \protect\circone\ for system sizes $L=4$, 6, and 8 at coupling $\xi=0.5$. (b) Collapse of the correlation function data in (a) with (2+1)d Ising critical exponents. (c) $S(\vec{Q})$ of Ising spins along path \protect\circtwo\ at coupling $\xi=1.5$. There is a first-order transition between two FM states at $h\sim2.8$, followed by a continuous FM-to-PM transition at $h_c=3.65$. (d) Collapse of the correlation function data in (c) close to $h_c$ with (2+1)d Ising critical exponents.}
\label{fig:sqvshmerge1}
\end{figure}

For $\xi/t\lesssim 1$,  the FM to PM phase transition is seemingly identical to that of the 2D TFIM, except for a small shift of the critical value of $h_c$. This statement is based on  the evaluation of  the correlation function of Ising spins, $S(\vec{Q})=\frac{1}{L^2} \sum_{bb'} \langle s^{z}_b s^{z}_{b'} \rangle e^{i\vec{Q}\cdot\vec{(r_b-r_{b'})}}$ with $b$, $b'$ running over all Ising spin sites and $\vec{Q}=(0,0)$. This data is shown in panels (a) and (b) of Fig.~\ref{fig:sqvshmerge1}. The quality of the data collapse with (2+1)d Ising critical exponents is very good, which suggests that the FM to PM transition along path \circone\ is still of (2+1)d Ising universality class~\cite{BatrouniScalettar2011}. 
Since the observed ferromagnetic ordering renormalizes the velocities, our result is consistent with the point of view that small velocity anisotropies that break Lorentz invariance are irrelevant \cite{Vojta00,Vafek02}.

When the velocity renormalization becomes sufficiently large, it can trigger the topological phase transition.
At $h =0$ and $\xi=1$, because of the Fermi surface nesting, the fermionic density of states diverges, as do spin and charge susceptibilities. At finite value of $h$, fluctuations of the Ising spins provide an interaction between the fermionic degrees of freedom and, owing to the Stoner instability, will potentially trigger an ordered state. Alternatively, a first-order transition can separate the two topologically distinct Dirac phases.

\begin{figure}
\includegraphics[width=\columnwidth]{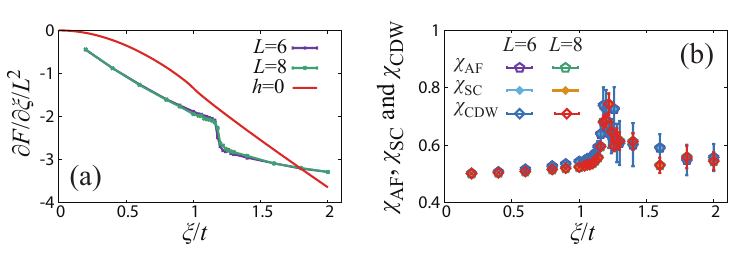}
\caption{Measurements along path \protect\circfive.
(a) Derivative of the free energy with coupling strength.
Here the $h=0$ case of $\partial F/ \partial \xi$ is also plotted and the value has been scaled down by a factor of two for comparison in the same plot. It shows a continuous transition. The jump for $h=2.0$ case ( data points with $L=6$ and $8$ ) directly show a first order transition around $\xi/t \simeq 1.2$.
(b) Fermionic spin-spin antiferromagnetic correlation functions $\chi_{\text{AF}}$, on-site s-wave pairing correlations $\chi_{\text{SC}}$, and CDW correlations $\chi_{\text{CDW}}$. These functions are degenerate and greatly enhanced in the vicinity of the $\xi/t$ value marked by the red dashed line in Fig.~\ref{fig:phase-diagram}.}
\label{fig:fsh2_0}
\end{figure}

To investigate the above scenarios, we compute $ \frac{1}{L^2}\frac{\partial F}{ \partial \xi}   =  \frac{1}{L^{2}}\sum_{j,\delta}\sgn(\delta) s_{j,j+\delta}^{z}\bigl\langle \bigl(c^{\dagger}_{j}c^{}_{j+\delta}\ e^{i\frac{\pi}{4}\sgn(\delta)}+\text{h.c.}\bigr)   \bigr\rangle$, the derivative of the free-energy density $F/L^{2}$ with respect to the control parameter $\xi$, along path \circfive. As shown in Fig.~\ref{fig:fsh2_0}(a), this derivative shows a clear jump at the transition point, which implies a first-order topological phase transition.

We have also calculated the fermionic correlation functions along path \circfive, such as the antiferromagnetic correlation function $\chi_{\text{AF}}=\frac{1}{L^{2}}\sum_{ij}(-1)^{i+j}\bigl(\langle s_{i}^{z}s_{j}^{z}\rangle -\langle s_{i}^{z}\rangle\langle s_{j}^{z}\rangle\bigr)$ with $s_i^z=\frac{1}{2}\bigl(c_{i\uparrow}^\dagger c^{}_{i\uparrow}-c_{i\downarrow}^\dagger c^{}_{i\downarrow}\bigr)$. Owing to the $\text{SU}_S(2) \otimes \text{SU}_\eta(2) \otimes Z_{2,\text{ph}}$ symmetry of the model, this correlation function is degenerate with CDW correlations $\chi_{\text{CDW}}=\frac{1}{L^2}\sum_{ij}(-1)^{i+j} \langle \eta^z_i \eta^z_j \rangle$ and on-site s-wave pairing correlations $\chi_{\text{SC}}=\frac{1}{2L^2}\sum_{ij} \langle \eta^+_i \eta^-_j \rangle$, built from bilinears $\eta^z_i=(c_{i\uparrow}^{\dagger}c_{i\uparrow}+c_{i\downarrow}^{\dagger} c_{i\downarrow}-1)/2$ and $\eta^-_i=c_{i\downarrow}c_{i\uparrow}$.   As shown in Fig.~\ref{fig:fsh2_0}(b), the various correlation functions are indeed degenerate, and all develop a peak at $\xi /t \simeq 1.2$, which is consistent with the relation $ \xi S(Q)/t L^2 = 1$ that describes the divergence of the density of states at the mean-field level.   Since there are no significant  size effects between the considered lattice sizes, the data is consistent with the absence of long-range order and associated breaking of the SO(4) symmetry.  We use a cross-hatched region in the Fig.~\ref{fig:phase-diagram} phase diagram to present an area where the SO(4) bosonic fluctuations are greatly enhanced. 
For the case of broken symmetry states, the coexistence of s-wave pairing and antiferromagnetism has been discussed in detail in Ref.~\onlinecite{Feldbacher2003,Assaad16}.   Note  that  at  the value of  $h$  considered in Fig.~\ref{fig:fsh2_0} (deep inside FM state), the Ising spins are only weakly fluctuating. The QMC simulation dynamics are thus slow (see SM~\cite{suppl}). 

\begin{figure}[htp!]
\includegraphics[width=0.85\columnwidth]{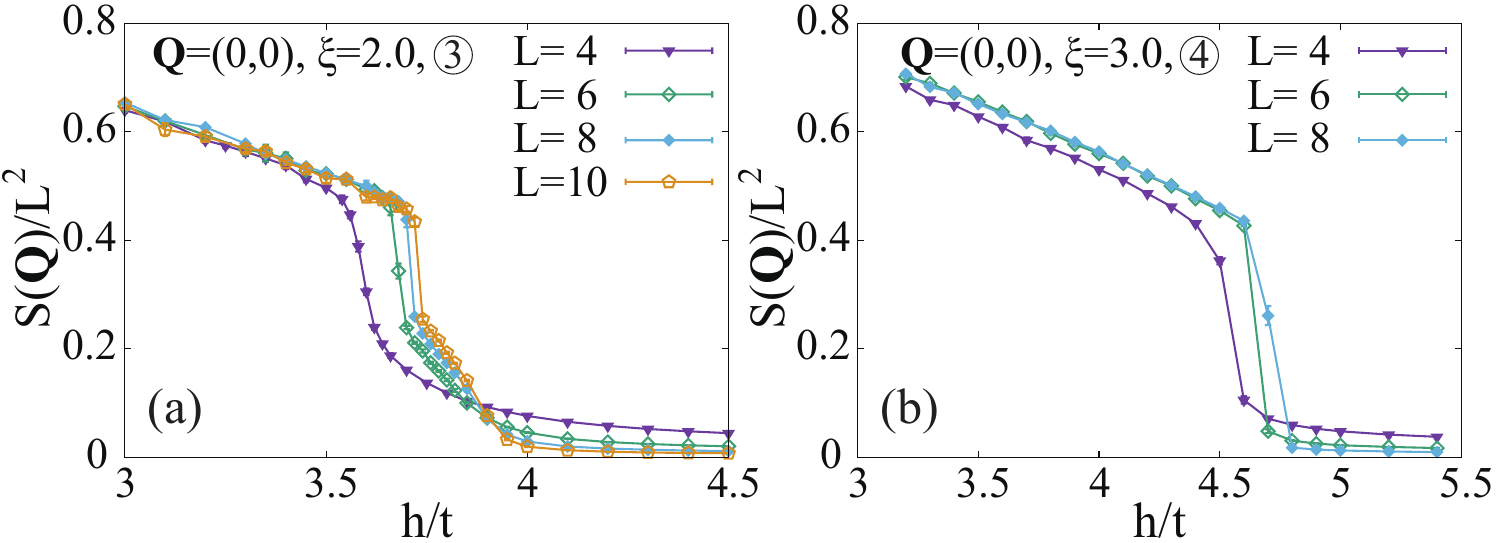}
\caption{(a) Correlation function $S(\vec{Q})$ of Ising spins along the path \protect\circthree\ for sizes $L=4$, 6, 8, and 10 with coupling $\xi = 2.0$. As a function of $h$, there is a first-order transition between two FM states, following by a continuous transition between FM and PM states. (b) $S(\vec{Q})$ of Ising spins along path \protect\circfour\ with coupling $\xi = 3.0$. There is a single first-order transition between FM and PM states.}
\label{fig:sqvshmerge2}
\end{figure}

At larger values of $h$, fluctuations of the Ising spins become stronger. Along paths \circtwo\ and especially \circthree\ and \circfour, we see first-order transitions between the Dirac phases with different topological index. As illustrated along path \circtwo\ in Fig.~\ref{fig:sqvshmerge1}(c), as a function of $h$, we first observe this first-order transition between the two FM states and then a (2+1)d Ising transition from FM to PM in the Ising spins, similar to what happens in Fig.~\ref{fig:sqvshmerge2}(a). For still larger $\xi$, the Ising spins undergo only a first-order transition from FM to PM states, as shown in Fig.~\ref{fig:sqvshmerge2}(b).

{\it Discussion}\,---\,We have investigated the robustness of Dirac fermions to fluctuations that break the $C_{4v}$ symmetry.  Using unbiased QMC simulations, we have shown that the Ising quantum phase transition remains in the (2+1)d Ising universality class, provided that the coupling between the Ising and fermion degrees of freedom is weak. This serves as a numerical proof that {\it small} velocity anisotropies in Dirac systems are irrelevant. At larger couplings, however, where the velocity is strongly renormalized, we observe enhanced spin and superconducting fluctuations around points in phase space where one of the velocities vanishes. The locking of antiferromagnetic correlations and s-wave superconductivity is a consequence of the SO(4) symmetry present in the model. At these points, the Berry phases of the Dirac cones interchange, and the transition turns out to be of first order.

Our model emphasizes fluctuations in velocity as opposed to fluctuations in the position of the Dirac points. It has some similarity to recent work reported in Ref.~\onlinecite{Schattner2015a}, and we can easily extend it to more complicated Fermi surfaces. The flexibility of our numerical approach allows for arbitrary couplings between Ising spins and Dirac fermions such that velocity and Dirac point fluctuations can be investigated.

{\it Acknowledgments}\,---\,The authors thank E.\ Berg and I.\ Herbut for helpful discussions. XYX and ZYM are supported by the National Natural Science Foundation of China (NSFC Grant Nos.\ 11421092 and 11574359) and the National Thousand-Young-Talents Program of China. KS is supported by the National
Science Foundation, under Grant No. PHY-1402971 at the University of Michigan and the Alfred P. Sloan Foundation. FFA is supported by the DFG under Grant FOR1807. KSDB and FFA gratefully acknowledge the hospitality of the Institute of Physics, Chinese Academy of Sciences. We thank the following institutions for allocation of CPU time: the Center for Quantum Simulation Sciences in the Institute of Physics, Chinese Academy of Sciences; the National Supercomputer Center in Tianjin; the John von Neumann Institute for Computing (NIC) for providing access to the supercomputer JURECA \cite{Jureca16} at J\"ulich Supercomputing Centre (JSC); and  the Gauss Centre for Supercomputing e.V. (www.gauss-centre.eu) for  providing access to the  GCS Supercomputer SuperMUC at Leibniz Supercomputing Centre (LRZ, www.lrz.de).

\appendix
\bibliography{main}

\end{document}